   \documentclass[preprint2]{aastex}

\shorttitle{LBV statistics}
\shortauthors{LBV statistics}

\begin{document}

\title{ LBV's and Statistical Inference }

\author{
  Kris Davidson\altaffilmark{1}, Roberta M. Humphreys\altaffilmark{1}, and 
  Kerstin Weis\altaffilmark{2}     }

\altaffiltext{1}
{Minnesota Institute for Astrophysics, 116 Church St SE, University of Minnesota
, Minneapolis, MN 55455; roberta@umn.edu} 

\altaffiltext{2}
{Astronomical Institute, Ruhr-Universitaet Bochum, Germany; 
kweis@astro.rub.de}

\begin{abstract}  

\citet{ST15} asserted that statistical tests disprove the standard view of 
LBVs, and proposed a complex alternative scenario.   But \citet{HWDG} showed 
that ST's test samples were mixtures of disparate classes of stars, and 
genuine LBVs statistically agree with with the standard view.  \citet{S16} 
objected at great length to this result.  Here we explain why each of 
his criticisms is incorrect.  We also comment on related claims made by 
\citet{ss16}.  {\it This topic illustrates the dangers of uncareful 
	statistical sampling and of unstated assumptions.}   

\end{abstract} 


      %%%%%%%%% ===-===   place marker

\section{The issue at hand}   %%% ===-===  SECTION 1   
\label{sec:intro}

  \noindent [{\,}This is a revised version of a paper that appeared earlier 
  in arXiv.   Here some of the arguments have been simplified and new remarks 
  have been added, especially in {\S}\ref{sec:gaia}, plus a brief Appendix 
  concerning the KS test.{\,}]  

  \vspace{2mm}

\citet{ST15} asserted that Luminous Blue Variable stars (LBVs or S Dor 
variables) have gained mass in binary systems.  This idea was motivated by a 
statistical claim that LBVs are spatially ``isolated'' from young objects 
such as O-type stars.  In the proposed scenario, each pre-LBV star has an 
extended lifetime due to the mass that it has gained, and an extra  velocity 
due to the ``kick'' (i.e., recoil) when its companion becomes a supernova.   
These two effects were said to explain the alleged isolation.    

    \vspace{1mm}  

But \citet{HWDG} noted some facts which invalidate the alleged isolation 
(``ST'' refers to Smith and Tombleson's paper):   
\begin{itemize}  
  \item{ About 1/3 of ST's Magellanic ``LBVs'' are not LBVs, and some of 
	  the others are doubtful. }  
  \item{ ST employed a statistical test that is not valid for mixed or 
	  contaminated sample sets. (See this paper's Appendix.) } 
  \item{ It has long been recognized that the less-luminous LBVs have 
    different evolutionary histories than the more luminous classical LBVs.  
    ST mixed the two classes indiscriminately. }    
  \item{ It turns out that the luminous classical LBV's {\it are\/} 
     statistically associated with O-type stars, as expected in the 
     in the standard view but not in the ST scenario. }  
  \item{ Lower-luminosity LBVs, consistent with their older expected  
     ages, are {\it not\/} statistically associated with O-type stars.  
     Hence the two classes do indeed differ, contrary to ST's implicit 
     assumption.}   
  \end{itemize} 
These results support the standard view of LBVs,  while the mass-gaining 
hypothesis fails the test that was cited as its motivation.  There is no 
evidence that LBVs are substantially older than expected.  Humphreys et al.\  
reviewed only the statistics and did not discuss failures in the physics 
of ST's scenario ({\S}\ref{sec:discuss} below).    \vspace{1mm} 

\citet{S16} has objected, at great length, to these conclusions.  Here we 
show that his critique has several disabling faults:     
 \begin{itemize} 
   \item{The main results in \citet{HWDG} are misquoted or misrepresented. } 
   \item{Smith modifies the original ST scenario and adds new assumptions. }  
   \item{Test criteria are likewise altered.  He attempts to transfer the 
	   burden of proof away from the mass-gaining hypothesis. }   
   \item{Crucial observational facts are not acknowledged. } 
   \item{Smith continues to emphasize the Kolmogorov-Smirnov (KS) 
           statistical test with faulty samples. } 
   \item{He offers no explanation for the obvious statistical difference  
           between high- and lower-luminosity LBVs. } 
 \end{itemize} 
Most of these flaws are evident if one consults the preceding papers, but    
the length of his critique may deter readers from doing that.  Moreover, 
Smith employs new assumptions and more KS tests which are invalid  
for specific reasons.  Meanwhile \citet{ss16} have introduced a new set 
of fallacies.  Hence the following account is necessary.  

   \vspace{1mm}  

In this paper ``ST,'' ``HWDG,'' and ``Smith'' refer to \citet{ST15}, 
\citet{HWDG}, and \citet{S16}.   ``HD94'' is  \citet{HD94}, which remains 
essential for this discussion despite its age.  (Smith often cites his own 
much later papers for concepts that were explained in HD94.)  
Quantity $D1$, the projected distance to the nearest O-type star, is 
statistically correlated with age, see ST and HWDG.  
Note, throughout, that ``LBV'' is essentially a synonym for ``S Dor 
variable.''  

  \vspace{2mm}

\section{Misconceptions?}   %%%  ===-===  SECTION 2 ?    

\label{sec:misconcepts} 

In his {\S}2, Smith identifies ``factual errors and misconceptions'' 
in HWDG;  but each example is either incorrect on his part, or 
inconsequential,  or requires  alterations of the ST model.  Hence 
we must review his list.  Since full explanations would be extremely 
tedious here, see the original papers for details.  The first two  
are most  crucial.  

  \vspace{1mm} 

(1) LBVs have higher $L/M$ ratios than other stars in the same parts 
of the HR diagram. Smith is simply wrong in calling this fact ``an 
assertion that has no empirical verification'' and ``a conjecture from 
single-star evolutionary models, not observations.''  
\citet{k89}, \citet{pp90}, \citet{s90}, \citet{s91}, \citet{v02}, and 
\citet{v12}  reported spectroscopic analyses that all showed abnormally 
low LBV masses compared to their luminosities.  Smith offers no reason 
to doubt them, but appears to demand a binary-orbit mass estimate -- which 
is practically unattainable because it would require LBVs in {\it eclipsing\/} 
double-line spectroscopic binaries.

In {\S}\ref{sec:discuss} below, we emphasize why 
$L/M$ is crucial for LBVs.  The mass $M$ may require a 
correction factor for rapid rotation,  but this obvious detail  
does not alter the basic reasoning. 

   \vspace{1mm}  

(2) Near the end of his {\S}2, Smith asserts,  as though it were well 
established, that ``statistically LBVs appear to receive an extra 
velocity spread (or longer lifetimes) beyond that given to the rest 
of massive stars.''  But {\it ST failed to justify that assertion,\/} 
see HWDG.  Then he demands formal tests ``rather than picking a few 
stars out of a sample.'' 
This view has two obvious faults.  First, HWDG did not merely 
``pick a few stars.''  They reviewed all 19 members of ST's   
Magellanic LBV sample and showed that seven of them definitely  
should not have been included, while five of the others are 
unconfirmed ``LBV candidates.'' They also reviewed all of the 
known LBVs in M31 and M33.    
Secondly,  the KS test that he advocates is {\it quite vulnerable to 
mixed samples and false sample members,\/} see this paper's Appendix.  

   \vspace{1mm} 

(3) A historical misrepresentation:  ``\citet{conti84} originally defined 
LBVs...''  In fact Conti originated the acronym LBV but not the class, 
and the term ``luminous blue variables'' appeared earlier 
\citep{sata74,hd79,hd84}.  Various authors later explored criteria to 
make LBVs or S Dor variables a meaningful species, e.g., \citet{lam86,
boh89,rmh89, w89};  and others in \citet{ld87} and \citet{dml89}.   
They led to the definition expressed in HD94.\footnote{    
   %%% FOOTNOTE
   In recent years many authors have used  relaxed definitions of  
   ``LBV,'' but those are physically almost useless, because they 
   include stars with differing structures and evolutionary states. }  

    \vspace{1mm}  

(4) Smith asserts that a star in the triple system  HD 5980 is a classical 
LBV, and its large $D1$ value would alter HWDG's statistical conclusions.  
This is untrue for three separate reasons. 
\ (a) As HWDG noted, its outburst did not match a normal LBV event; so we 
cannot include it in the ``confirmed'' sample set.  It may have been a 
different type of eruption caused by a close companion. 
\ (b) One of its companions is an O-type star \citep{hd5980}.  Therefore, 
if it is an LBV, then it contradicts ST's ``isolation'' and extended-age claims. 
\ (c) Even if we ignore the O-type companion, HD 5980 has a smaller $D1$ 
value than any star in HWDG's set LBV2, the Magellanic lower-luminosity 
LBVs. This fact would slightly strengthen HWDG's conclusions, not weaken them.         

   \vspace{1mm}  

(5) He next affirms at length that MWC 112 is different from 
Sk~$-69^\circ$ 142a; but  HWDG did not say otherwise.\footnote{    
    %% FOOTNOTE
    A preprint version repeated the old (c.1990) confusion between 
    these two stars, but it was corrected before formal publication.  
    This had no effect on any conclusion. }   

    \vspace{1mm} 

(6) According to Smith, ``another misconception expressed by HWDG'' 
concerns LBV velocities.  His explanation adds new assumptions and 
complications to the ST scenario, see {\S}\ref{sec:vels} below.   

     \vspace{1mm}  

(7) Smith notes that the ST model allows LBVs in binary or multiple 
systems, a possibility not clearly stated in their original paper.  
(They emphasized the alleged ``isolation'' of LBVs and ``kicked'' 
velocities.)  Since most of his comments about binaries are also 
consistent with the standard view of LBVs, they provide no support 
for the mass-gaining hypothesis.    

     \vspace{1mm}  

(8) In his $\S$3, he employs the KS test to assert that ``candidate LBVs'' 
should be included in the statistical sample.  The same reasoning would 
show that grapes can be included in a sample of cherries, based on their 
similar size distributions relative to oranges.  By quoting p-values, 
dispersions, and other formal concepts, one can make this fruity conclusion 
seem palatable --  unless we examine the objects.  In the LBV case, a 
particular set of ``candidate'' stars have a spatial distribution that 
is similar to LBVs in some respects, hence Smith deems them to be LBVs 
for statistical purposes.  Worse, about 35\% of the sample members are 
almost certainly not LBVs (see HWDG) -- but he still includes them 
without explaining why.  Remarks by \citet{nmgb}, concerning p-hacking 
and related techniques, are pertinent to this as well as other  aspects 
of the discussion.

\section{Cherry-picking}  %% ===-===   
\label{sec:cherryp}

As noted above, the ST sample of ``LBVs'' was a mixed set of objects, 
fundamentally unsuitable for a KS test.  
But Smith's {\S}4 is titled ``cherry-picking the sample,'' 
with an implication that subdividing it would amount to fudging the 
statistics.   Astronomers usually refer to this activity as stellar 
classification, not cherry-picking, and the stars' physical differences 
are real.

   \vspace{1mm}  

(1)  Referring to HWDG's  set LBV1 of three classical LBVs in the LMC,  
Smith objects:  ``One may question the validity of selecting the tail end 
of a distribution...,'' and ``if one extracts the tail end...''  He means 
that these stars have the smallest values of the  distribution parameter 
$D1$.  But $D1$  played no part in selecting them!  Luminosity was the 
criterion, and {\it set LBV1 was recognized long before anyone evaluated 
the $D1$ values,\/} see HD94.  In ST's mass-gaining scenario there is no 
evident reason for $D1$ to be correlated with luminosity.  But in the standard 
view it's obvious:  classical LBVs are younger than the less-luminous LBVs.  
What Smith calls ``the tail end'' is a physically distinct set. 

  \vspace{1mm}  

(2) He alleges an ``unquantified selection bias'' in set LBV1.  But those 
three stars were identified more than two decades ago as the classical LBVs 
in the Magellanic Clouds (HD94). No more have been found since.
(If HD 5980 is an exception, then it would strengthen HWDG's formal results
as noted in {\S}\ref{sec:misconcepts} above.)   Every member of the 
less-luminous set LBV2 is definitely below the luminosity criterion.  Thus 
it is hard to see what Smith means by ``selection bias'' for such a tiny, 
well-defined set with no borderline members.  

   \vspace{1mm}  

(3) Next, Smith misrepresents a crucial result.  He says 
that HWDG ``found that [\,set LBV1 has\,] a similar distribution ...  to 
late O-type dwarfs.''  That should not be true in the standard model, 
because the high-$D1$ tail of the O8-O9 distribution (Fig.\ 1) presumably 
represents stars that are older than classical LBVs should be.  In fact, 
however,  the term ``similar distribution'' is absurd for a sample of only 
three objects.  As Figure 1 shows, late-O stars with $D1 > 20$ pc are 
obviously irrelevant to set LBV1 which has $D1 < 15$ pc!   HWDG's result 
was much simpler:  since the stars in set LBV1 have fairly small values of 
$D1$, {\it one cannot prove that they differ from the O-star distribution.\/}  
At first sight this conclusion appears very weak, but it is important because 
it would not be true in ST's model.

All three stars in set LBV1 have $D1$ in the range 4--12 pc, close to the 
median for O6-O9 -- i.e., the median for all O-type stars that can be 
3 to 4 Myr old.  Classical LBVs have such ages in the standard view, see 
item 5 below.

   %%%%  ===-=== 

    \begin{figure}[ht] 
    \figurenum{1}
   % \epsscale{0.5}
    \plotone{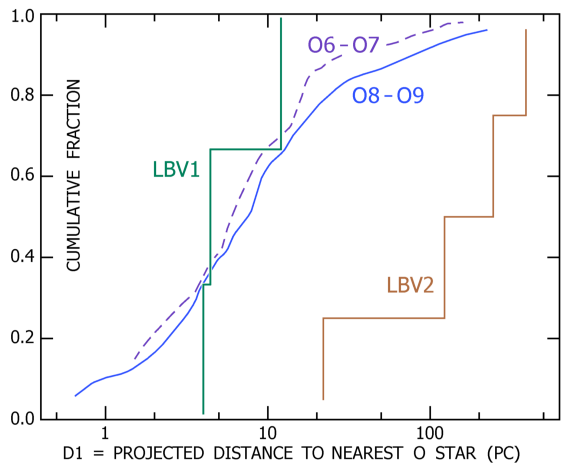}
    \caption{Distributions in the Magellanic Clouds, plotted in the 
       style used for KS tests.  Sets LBV1 and LBV2, the classical 
       and less-luminous LBVs, were recognized long before $D1$ was 
       evaluated for any of them (see HD94 and HWDG) .}   
    \end{figure}

    \vspace{1mm}

(4) Regarding the older, less luminous LBVs, Smith belatedly states 
that the ST comparison sample of red supergiants must now be 
revised by deleting the lowest-mass members.  Instead of presenting 
the new distribution, he {\it conjectures\/} that it would alter 
HWDG's results.

Instead of arguing about RSG comparison samples, we can simply examine 
the four stars in HWDG's ``set LBV2,'' the confirmed less-luminous LBVs 
in the Magellanic Clouds.  Two of them, R71 and R110, are located more 
than 200 pc from their nearest known O-type stars.  Does this fact 
imply greatly extended lifetimes?  Probably not, since the expected 
ages of lower-luminosity LBVs appear adequate to account for it.
In the standard scenario outlined by HD94 and HWDG, these objects 
should be more than 5 Myr old, exceeding all normal O-type stars;  
see \citet{ek12}, \cite{ch15}, HWDG's Figure 1, and the 5-Myr isochrone 
in Fig.\ 12 of \citet{m05}.\footnote{  
    %%% FOOTNOTE    
    Smith quotes lifetimes of 10 Myr for stars of 20 $M_\odot$, 
    but they evolve beyond O-type in half that time. }  
Therefore, we can reasonably surmise  that massive stars which formed 
along with R71 and R110 have now evolved beyond O9.  Rotation and 
extra mass loss may reinforce this explanation.  In order to argue  
that these stars require extended lifetimes in the sense that ST 
proposed -- a strong assertion -- one would need to disprove 
the above assessment, not merely express hypothetical reasons why 
it might be wrong. 

    \vspace{1mm}  

(5) Smith next asserts that the standard view requires classical LBVs to 
be associated with O2-O5 stars, rather than later-type O stars.  This claim 
is illogical, since {\it LBVs are expected to be older than the O2-O5 stars.\/}  
Classical LBVs should be roughly 3.5 Myr old in the standard 
view, while the maximum for a normal O5 star is about 2.5 Myr and, more 
important, the {\it median\/} for O2-O5 stars is less than 1.5 Myr.  
These estimates are based on \citet{ek12}, \citet{ch15} with LMC 
metallicities, and the locations of R127 and similar stars in the HR 
diagram.  Classical LBVs might even be older than 4 Myr if they rotate 
rapidly.    

In Smith's Figure 2, the O2-O5 distribution greatly differs from O6-O9  
because 40\% of its members have separation parameters $D1 < 3$ pc.  
Such small values indicate average ages less than 1 Myr, too young 
for us to expect a spatial correlation with LBVs.  The O6-O9 stars 
provide a far more logical comparison, since many of them have ages 
of 3--4 Myr.  In summary, his Figure 2 relates to star formation 
history, not LBV physics.  

    \vspace{5mm} 

\section{Velocities}   %% ===-===  
\label{sec:vels} 

HWDG noted that LBVs do not have the high velocities that would occur 
in the scenario that ST described.  In order to transfer a significant 
amount of material, the assumed type of binary has to be close, 
with orbital speeds of the order of 200 km s$^{-1}$.  Allowing for 
complications, one can reasonably expect runaway velocities well 
above 50 km s$^{-1}$ in at least a few cases, contrary to observations. 

  \vspace{1mm}  

Smith now explains this by declaring, after lengthy preliminaries, 
that the required supernova occurs only after its star has been reduced 
to about 4 $M_\odot$ -- so the other (future LBV) star's orbital 
speed becomes less than 30 km s$^{-1}$.  But he does not say why this 
should always be true. ST employed the word ``kick'' repeatedly, in a 
way that implied a major role for enhanced velocities.  Therefore 
Smith's reassessment alters their model in order to cancel a failed 
prediction.\footnote{ 
    %%% FOOTNOTE 
    Incidentally, we note that the recoil speed would be 
    small if mass transfer occurs when the primary star is a 
    cool supergiant, i.e., if the orbit is larger.  There are  
    enough free parameters to enable various conjectures 
    like this. } 

  \vspace{1mm}   

His {\S}5 recounts more KS tests, now referring to velocities 
in M31 and M33.  They produce no significant result apart from 
the obvious lack of high-speed LBVs.  

  \vspace{1mm}

Then he adjusts a model of recoil speeds to optimize its 
KS $p$ value, and finds that it differs from the 
zero-extra-velocity case.  But this result is inevitable 
because the parameter of interest is non-negative, 
$V_k \geq 0$.  Even if its true value is zero, statistical 
errors always repel the formal best-fit value away from  
$V_k = 0$ because that is the limit of the allowed range.  

   \vspace{1mm}  

One particular misconception should be noted here to forestall it 
from propagating   
in the literature.  Smith says that R71 ``has a radial velocity that is 
offset by $-71$ km s$^{-1}$ relative to the systemic velocity of the 
LMC.''  This difference is meaningless, because the  star's velocity 
agrees well with the rotation curve {\it at its particular location.\/}  
As HWDG carefully stated,  the LMC LBV velocities are consistent ``with 
their positions based on the H I rotation curve available at NED.''\footnote{ 
   %% FOOTNOTE 
   https://ned.ipac.caltech.edu/level5/Sept05/Sofue/Sofue7.html }

\section{AG Car and Gaia} 
\label{sec:gaia}  

\citet{ss16} recently noted that preliminary Gaia data appear to place 
AG Car only about 2 kpc from us, rather than 6 kpc as expected.  Since 
this star is often cited as a classical LBV, they speculate that further 
data will lead to a revolution in this topic and a vindication of the 
ST scenario.  That view has some obvious difficulties. 

  \vspace{1mm}

First, though, they misrepresent the state of the topic in general, by 
asserting that ``several aspects of [the] traditional view have started to 
unravel.'' This is a serious exaggeration.  Some of the examples 
that they cite to support it concern superficial details rather 
than fundamentals,  while the others have little to do with the basic LBV 
problem. Two of their examples  -- one involving $\eta$ Car, and 
the other repeating ST's claims -- are flatly erroneous.  Smith and 
Stassun refrain from citing papers that disproved their arguments -- 
e.g., \citet{dh12} and HWDG.  

   \vspace{1mm}

Concerning AG Car:  The quoted Gaia parallax has a stated $1\sigma$ uncertainty 
worse than $\pm \, 50$\%.  This is bad enough, but in addition, since it 
is far worse than the expected capability of the instrument, we can 
reasonably suspect that major systematic errors have not yet been diagnosed.  
Distances of classical LBVs in M31, M33, and the LMC are far more 
robust, and they all support 
the standard view (see HWDG).  Given their mutual consistency, and 
the provisional nature of early Gaia results, it is highly 
imprudent to claim that $D > 4$ kpc has been ruled out for AG Car.

Another objection concerns physics.  The methods used to estimate 
$L/M$ from spectra (see refs.\ in {\S}\ref{sec:misconcepts} above) are 
not very sensitive to distance.  If we conceptually move AG Car to 
$D <  2.5$ kpc,  then it has a truly peculiar set of attributes:     
(1) $L \gtrsim 10^5 \; L_\odot$, but (2) $M \sim 5 \; M_\odot$    
which is amazingly small for that luminosity, while (3) it has 
hydrogen and (4) there's no evident companion.  Meanwhile (5) it mimicks 
the spectra and behavior of classical LBVs in M31, M33, and the LMC!  
Frankly, this combination lacks credibility unless some more concrete evidence 
appears.

Smith and Stassun speculate that all these items might be explained 
in a binary context, but their story is entirely qualitative with  
ambiguous choices at every juncture.  In most respects  it scarcely 
resembles the  \citet{ST15} scenario;  AG Car is even said to belong 
to the Car OB1/OB2 association, ironic in light of ST's ``isolation''  
argument.   Rather than offering any clear insights into the LBV 
phenomenon, this view makes AG Car a strange object without  
genuine counterparts.

Considering all the known facts, the 
{\it simplest\/} (and not very surprising) explanation is that the 
Gaia parallax and proper motion contain one or more systematic 
errors that have not yet been diagnosed at this early stage.  Smith 
and Stassun also discuss three other objects that are not worth 
reviewing here.   The significance of $L/M$ is emphasized below.

\section{Discussion}   %% ===-===  
\label{sec:discuss} 

    \vspace{1mm}   

The main value of this dispute does not concern the scenario proposed by 
\citet{ST15}, which is not supported by the statistical tests when careful 
sample sets are used.  Instead, {\it this case illustrates two or 
three famous but often neglected principles of scentific evidence.}    

   \vspace{1mm}  

First, though, concerning only LBVs, the crucial point is this:  
\citet{HWDG} showed that bright classical LBVs {\it are\/} spatially 
correlated with O-type stars while the less luminous LBVs {\it are not.\/} 
This double fact was not previously noticed, it indicates an age difference 
consistent with the standard account of LBVs, and it was not expected 
in the ST scenario.  In other words it is unforeseen evidence for  
the standard view.  \citet{S16} later objected to many details, but 
did not offer any concrete reason to doubt this result.  

    \vspace{1mm} 

Regarding broader principles, this case highlights the dangers of 
statistical formalism applied to defective samples.   As illustrated 
in the Appendix below,   the Kolmogorov-Smirnov test  
gives false results if the sample set is a mixture of 
disparate classes.   ST's sample of Magellanic ``LBVs'' contained 
at least three physically distinct classes of stars, and therefore 
produced illusory $p$ values.  If a member of a statistical sample 
appears doubtful for some identifiable reason, {\it then that member 
should be removed.\/}   This policy would have averted serious errors 
in Smith and Tombleson's arguments.  

    \vspace{1mm}   

Another fundamental principle concerns burden of proof.  Here we have two 
competing scenarios: 
\begin{itemize} 
\item{ The conventional view of LBVs has been familiar since the 1980s.     
Essentially it consists of just one hypothesis, that the observed 
LBV events result from instabilities that arise 
when $L/M$ exceeds about half the Eddington Limit (HD94; \citealt{v12,kd16}). 
This has not been proven yet, but it is conceptually straightforward 
and theoretically credible,  it appears consistent with existing data and 
with existing evolution models, and no one has identified any errors 
serious enough to require a major change in the concept. }  
\item{ \citet{ST15} propose a different scenario motivated by statistical 
arguments.   It is undeniably more complicated,  it has multiple free 
parameters,  it requires many assumptions that the authors did not 
clearly acknowledge, and it offers no physical rationale for the LBV 
instability. }  
\end{itemize} 
According to normal standards the larger burden of proof rests on Smith 
and Tombleson, because their model entails more hypotheses and leaves 
more facts unexplained.  They appear to assume that a statistical 
failure of the standard model constitutes evidence in favor of theirs. 
This is illogical even if one discards the standard model, because 
other, simpler possibilities exist.  (For instance, rapid rotation can extend 
the lifetime of an LBV progenitor.)  Moreover, HWDG showed that the 
standard view does not fail the tests that ST proposed.  Nevertheless, 
\citet{S16} attempted to shift the burden of proof onto the conventional 
model by demanding  additional tests with new comparison sets, dynamical  
proof of the $L/M$ ratios, etc.   None of those objections constitutes 
positive evidence for the mass-gaining hypothesis.     

   \vspace{1mm}  

Now consider physics instead of statistics.  Any viable theory of LBVs 
must explain three conspicuous traits.  
  \begin{itemize} 
  \item{ These objects exhibit high-mass-loss episodes, or at least 
           expanded-radius events, with particular characteristics 
	   described in HD94 and elsewhere. } 
  \item{ They lie near a locus in the HR diagram, often called the LBV 
	  instability strip. However, most stars in that strip 
          are not LBVs. } 
  \item{ They have $L/M \sim 0.5 \, (L/M)_\mathrm{Edd}$, anomalously  
	  large for their locations in the HR diagram.\footnote{   
             %%% FOOTNOTE 
         	  As mentioned in {\S}\ref{sec:misconcepts}, rapid  
		  rotation might entail a lessened ``effective mass.'' }   
          Non-LBVs in the instability strip have larger masses and 
	  therefore smaller $L/M$. } 
  \end{itemize} 
The standard view accounts for these facts as described in  HD94 
and \citet{v12}.  Smith and Tombleson, however, 
did not give any clear reasons for them.

If we adapt the ST scenario by 
invoking the same instabilities as the standard model, then the   
hypothesis becomes essentially this: a star gains mass from its 
companion, and {\it after that it evolves in the same way as the conventional 
view of LBVs.} \,  But why is the mass exchange necessary?  If LBV progenitors 
need extended lifetimes, rapid rotation is arguably a simpler alternative.   
And, as explained above and in HWDG, there is no clear evidence 
for substantially extended lifetimes.  We suspect that the mass-gainer 
hypothesis will soon be revised to ``maybe some LBVs'' instead of 
ST's ``certainly all of them.''   

    \vspace{1mm}  

The concept of just two classes of LBVs is very likely an  
idealization of the truth,   
because we cannot easily guess what happens in borderline cases, and, 
besides, rotation and other parameters surely complicate the physics.  
Perhaps LBVs form a continuous mass- and rotation-dependent set, and 
in almost every 
case the instability arises at the time when $L/M$ first exceeds some 
critical value.  For the less-luminous LBVs, that occurs after a 
complicated series of evolutionary stages.  These thoughts seem adequate 
to explain the instability strip, even if  the truth is somewhat more 
complex.  

  \vspace{1mm}   

There is nothing wrong with hoping that binary interactions play some 
interesting role in this story, but it is wrong to {\it assume\/} 
that they do.  In order to replace the long-standing scenario
with something very different, one needs to satisfy the 
classic Hume-Laplace-Truzzi-Sagan rule:  extraordinary claims 
require extraordinary evidence. 

  \vspace{2mm} 

---  ---  ---  ---  ---  ---  ---  ---  ---  ---  ---  ---  

  \vspace{2mm}

\noindent{APPENDIX:  {\it Why the KS test needs careful sampling} }  

   \vspace{1mm}  

Imagine two physically distinct classes of objects with some distribution 
parameter $x$.    Suppose
that both classes, \underline{A} and \underline{B}, have accurately known 
distribution functions 
$f_\mathrm{A}(x)$ and $f_\mathrm{B}(x)$.
In Figure 2 these are plotted in the style used for a 
KS test.\footnote{   
  %%% FOOTNOTE 
  Expressed in terms of $\, u = \log_{10} x$, distributions A and B   
  are Gaussian with $(u_0, \, \sigma_u)$ = (1.0, 0.36) 
  and (1.6, 0.24) respectively.  Sample set C was produced by a  
  random simulation. }
Next suppose that we are given 
a list of $x$-values for 24 unfamiliar objects 
called ``type C,''  also shown in the figure.  Formally the KS test shows 
that they represent a new, intermediate class different from both A and B.   
The $p$ value is about 0.03 relative to B and even smaller for A.  

    \vspace{1mm} 

   %%%%  ===-=== 
     \begin{figure}[ht] 
     \figurenum{2}
   % \epsscale{0.8}
     \plotone{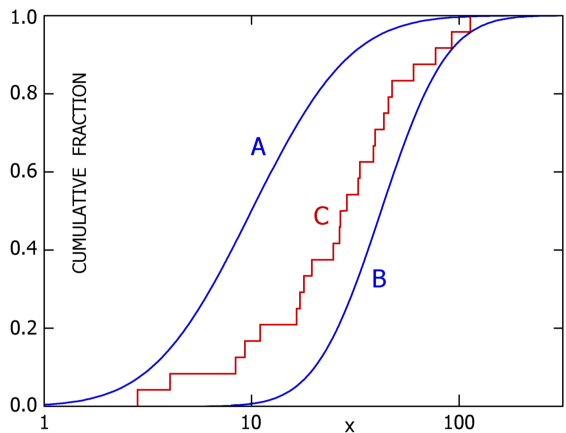}
    \caption{Distributions in a synthetic example.  Set C comprises 
             24 objects that have not been examined closely, apart 
             from their $x \,$-values. }    
     \end{figure}

However  -- as the reader may have guessed from the context -- this 
conclusion is flatly wrong, because distribution C is really a mixture 
of 9 random A's and 15 random B's.  (In order of increasing $x$, objects 
1--5, 7, 9, 12, and 14 belong to class A.)  Evidently the KS test 
{\it gives meaningless results\/} in a case like this with a mixed 
sample set.  

    \vspace{1mm}  

This elementary example closely mirrors the test that \citet{ST15} 
invoked in their {\S}2.2.  
Their indirect comparison sets (O-type stars, RSGs, etc.) do not affect 
the issue.   In the same manner as class C formally differed 
from both A and B in our example, Smith and Tombleson concluded 
that LBVs are statistically uncorrelated with particular classes of stars. 
(See their Figure 4.)    That was incorrect because,  
as HWDG showed, ST's sample set was a mixture of classical LBVs, older 
less-luminous LBVs, and non-LBVs.  The only way to justify that KS test 
is to insist that their sample was acceptable despite contrary evidence. 

  \vspace{2mm}  

  ---  ---  ---  

  \vspace{2mm}

\acknowledgements
We thank Dominik Bomans, Michael S.\ Gordon, and John C.\ Martin 
for useful comments.

%%%%%%%%%%%%%  ===-===  

\end{document}